\documentclass{JenniferZonneveld}

\def\Journal#1#2#3#4{{#1} {\bf #2}, #3 (#4)}

\def\PRD{{\em Phys. Rev.} D}

\def\be{\begin{equation}}
\def\ee{\end{equation}}
\def\bea{\begin{eqnarray}}
\def\eea{\end{eqnarray}}

\newcommand{\Photo}{}

\begin{document}
\vspace*{4cm}
\title{SEARCH FOR {\it CP} VIOLATION IN $B^0_s$ MIXING AT LHCB}

\author{ J.B. ZONNEVELD }

\address{University of Edinburgh, Particle Physics Experiment, Old College,\\
South Bridge, Edinburgh EH8 9YL\\jennifer.brigitta.zonneveld@cern.ch}

\maketitle\abstracts{
One of the main goals of the LHCb experiment is to determine the mixing-induced {\it CP} violating phase $\phi_s$
in $b \rightarrow c\bar{c}s$ transitions. Assuming the Standard Model, $\phi_s$ is precisely predicted, hence
new physics could easily affect the measurement. The most precise single measurement using
$B^0_s \rightarrow J/\psi K^+K^-$ data collected in 2015 and 2016 is presented along with a total combined
value including all LHCb $\phi_s^{c\bar{c}s}$ analyses. The result is further combined with Tevatron
and other LHC experimental results.}

\section{Introduction}

\begin{figure}[h]
\begin{minipage}{0.49\linewidth}
\centerline{\includegraphics[scale=0.55,trim={0cm 18cm 0cm 2cm},clip]{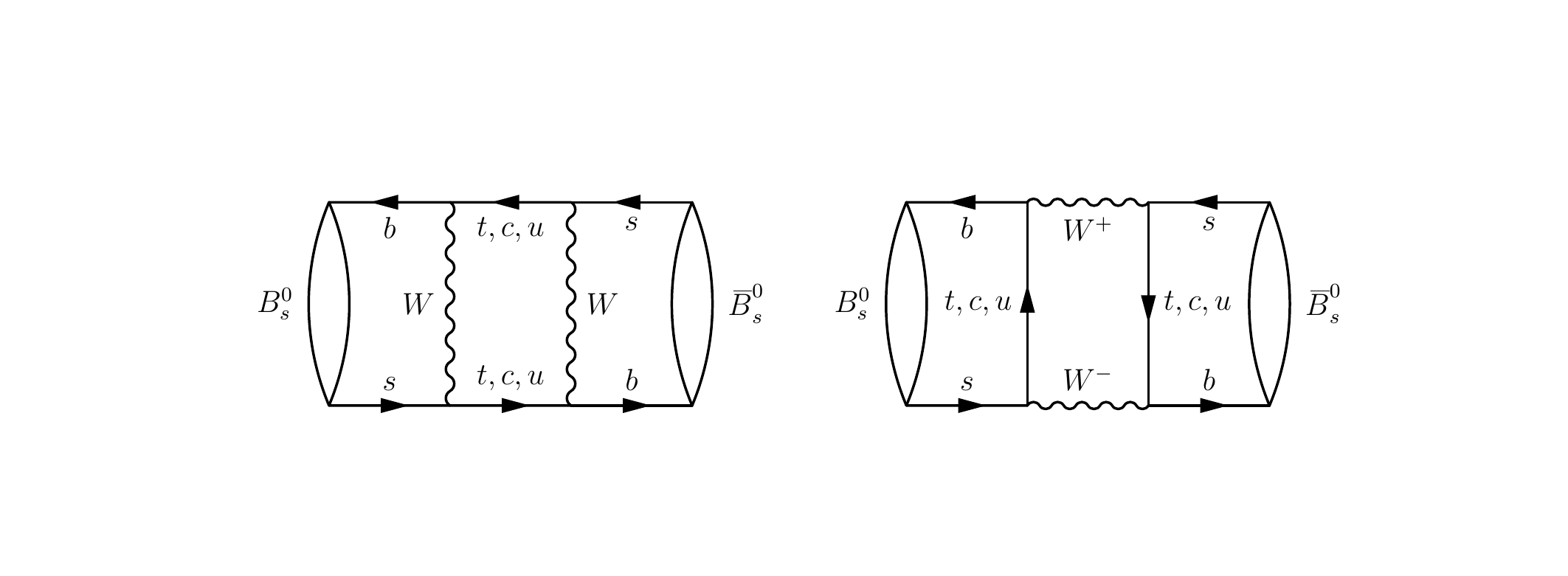}}
\end{minipage}
\hfill
\begin{minipage}{0.49\linewidth}
\centerline{\includegraphics[scale=0.55,trim={0cm 18cm 0cm 4cm},clip]{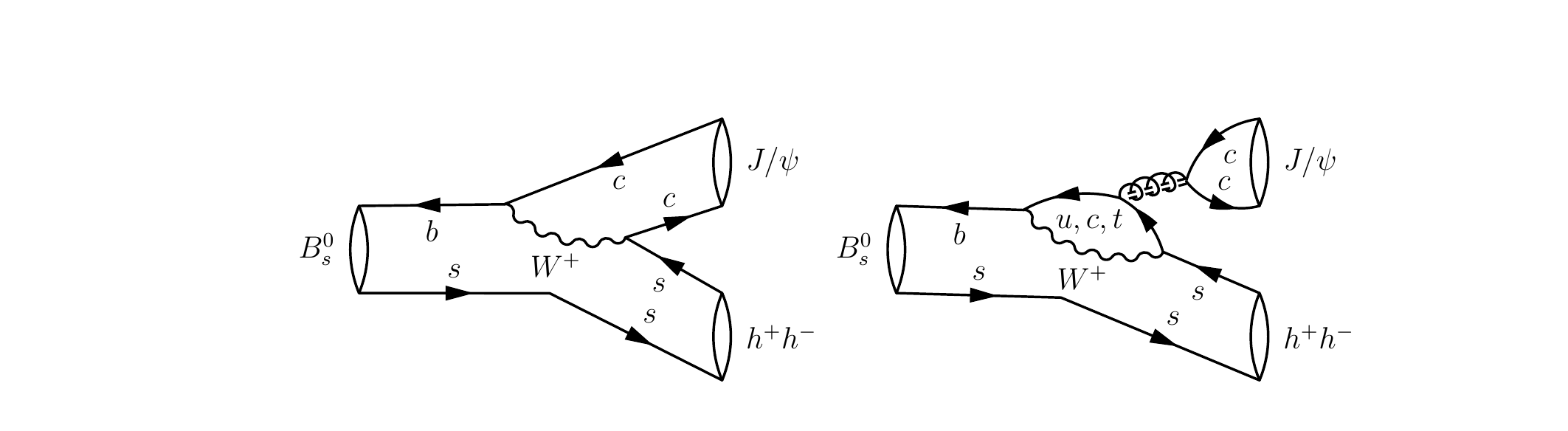}}
\end{minipage}
\caption[]{Feynman diagrams of $B^0_s - \bar{B}^0_s$ mixing (left), and SM tree and penguin contributions to the $B^0_s \rightarrow J/\psi h^+h^-$ decay, 
	where $h = \pi, K$ (right).}
\label{fig:decay}
\end{figure}

Interference between decay of $B^0_s$ mesons into {\it CP} eigenstates through $b \rightarrow c\bar{c}s$ transitions directly
and via $B^0_s - \bar{B}^0_s$ mixing gives rise to the {\it CP} violating phase $\phi_s$. Assuming the Standard Model (SM), and ignoring
sub-leading higher-order contributions (Figure~\ref{fig:decay}), $\phi_s$ can be related to $-2\beta_s$ as the following~\cite{CPtheory}:
\begin{equation}
	\phi_s = \phi_M - 2\phi_D = -2 \beta_s = -2 arg(\frac{V_{ts}V^*_{tb}}{V_{cs}V^*_{cb}}),
\end{equation}
where $\phi_M$ is the mixing phase and $\phi_D$ the decay phase. Global fits to experimental data lead to a very precise prediction
of $-2\beta_s = -36.9^{+1.0}_{-0.7}$~mrad~\cite{ckm}. If non-SM particles would play a role in the $B^0_s - \bar{B}^0_s$ oscillation, they could
significantly affect the measured value, which makes $\phi_s$ very interesting to study.

\section{Run 2 analysis of $B^0_s \rightarrow J/\psi K^+K^-$}

The most recent $\phi_s^{c\bar{c}s}$ measurement performed at LHCb is the time-dependent angular analysis of $B^0_s \rightarrow J/\psi(\rightarrow \mu^+\mu^-) 
\phi(\rightarrow K^+K^-)$. The decay has the advantages of a large branching fraction and large reconstruction efficiency. The main background contribution, originating
from combinatorial events, is suppressed by a Boosted Decision Tree~\cite{bdt} that is trained while avoiding variables that could introduce a decay time bias. 
The combined 2015 and 2016 data sample, which corresponds to a total integraded luminosity of $1.9\ \textrm{fb}^{-1}$ at a centre-of-mass energy of $\sqrt{s}$ = 13 
TeV, contains $117694 \pm 364$ signal candidates. 

\begin{figure}[h]
\center
\includegraphics[scale=0.6,trim={0cm 22cm 0cm 2cm},clip]{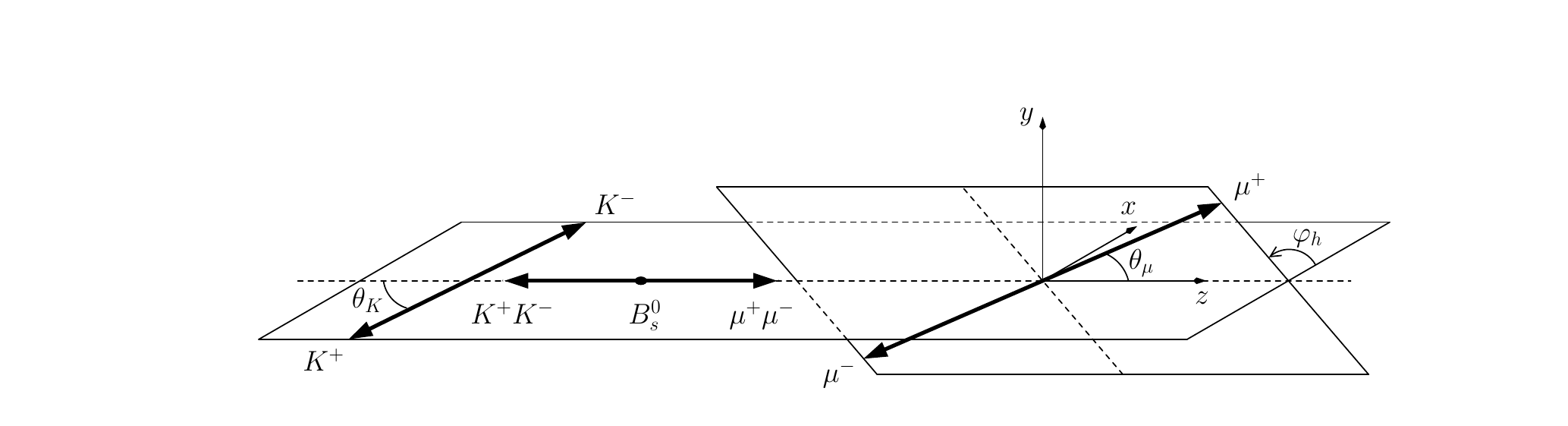}
	\caption[]{The angular distribution of $B^0_s \rightarrow J/\psi(\rightarrow \mu^+\mu^-)\phi(\rightarrow K^+K^-)$~\cite{hel}. The angle between 
	$K^+$ momentum in $\phi$ rest frame and $\phi$ momentum in $B^0_s$ rest frame is defined as $\theta_K$; the angle between $\mu^+$ momentum in 
	$J/\psi$ rest frame and $J/\psi$ momentum in $B^0_s$ rest frame is defined as $\theta_\mu$; the angle between $J/\psi$ and $\phi$ decay plane is defined as 
	$\phi_h$.}
\label{fig:helicity}
\end{figure}

Due to angular momentum conservation the final state is an admixture of {\it CP}-even and {\it CP}-odd components, with orbital angular momentum of $L=0, 2$ and $L=1$, 
respectively. The kaons in the three polarisation states originate mainly from the decay of the $\phi(1020)$ resonance (P-wave), however there is also a {\it CP}-odd 
contribution coming from an S-wave $K^+K^-$ ($\sim2\%$). An analysis of the decay angles of the kaons and muons is required to disentangle the even and odd components 
and determine $\phi_s$. Three helicity angles are defined as in Figure~\ref{fig:helicity}.

\begin{figure}[h]
\center
\includegraphics[scale=0.4]{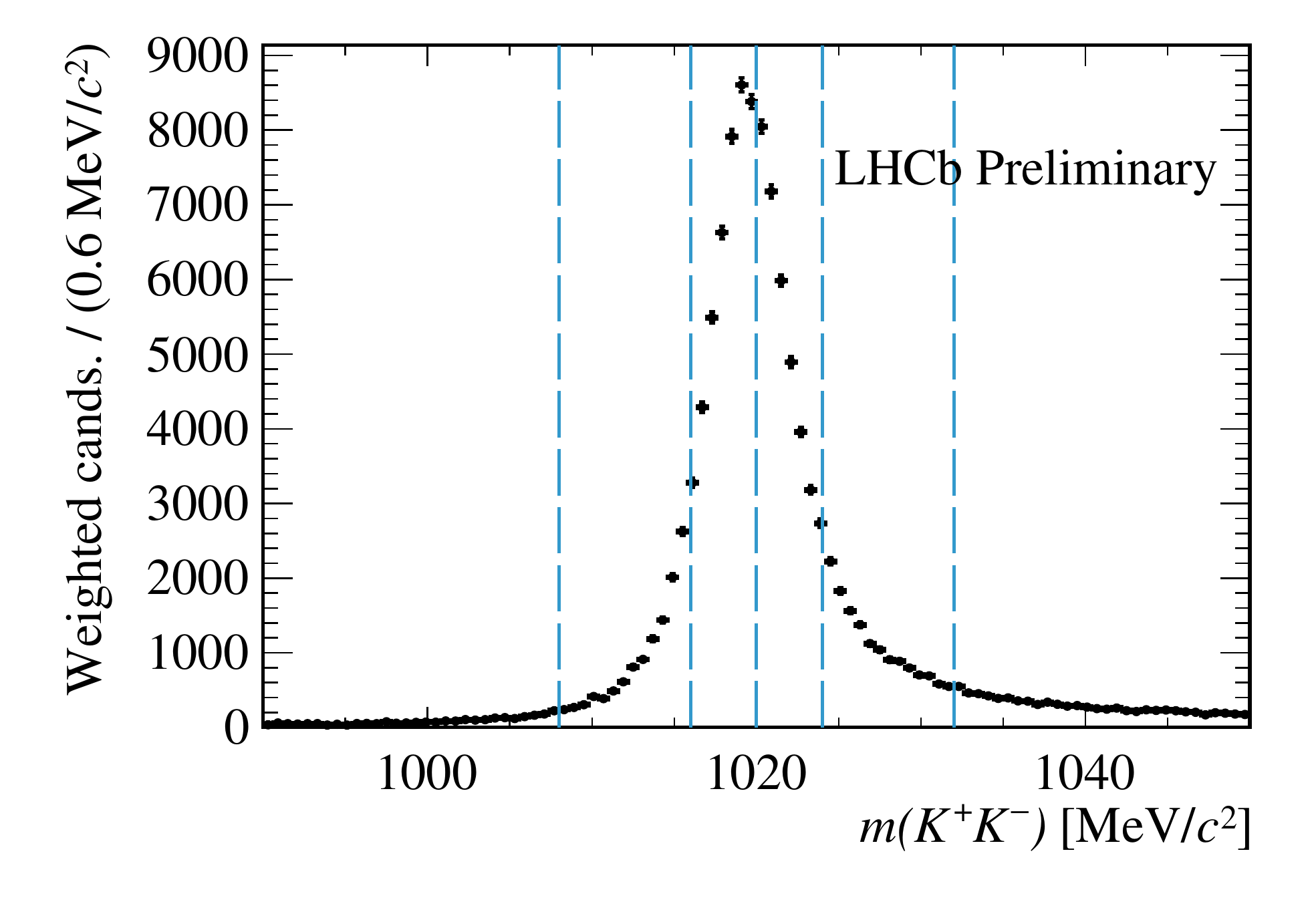}
	\caption[]{Distribution of the $K^+K^-$ invariant mass from selected $B^0_s \rightarrow J/\psi K^+K^-$ decays~\cite{phis}. The six $m(K^+K^-)$ bins 
	[990, 1008, 1016, 1020, 1032, 1050] are indicated by vertical blue dashed lines.}
\label{fig:mkk}
\end{figure}

To account for the S-wave, the data sample is split into six $m(K^+K^-)$ bins ([990, 1008, 1016, 1020, 1032, 1050] MeV/$c^2$, see Figure~\ref{fig:mkk}).
In each bin the interference between the P- and S-wave contribution is expressed in an effective coupling factor, $C_{SP}$. This is determined from 
simulation by integrating the interference in each bin. The flat broad S-wave contribution is considered to be an $f_0(980)$ resonance. The final fit is performed by 
simultaneously fitting to each $m(K^+K^-)$ bin with a floating fraction $F_S$ as $C_{SP}*F_{S}$. The background is removed by applying the 
\textit{sPlot}~\cite{sPlot} technique.

The $B^0_s - \bar{B}^0_s$ oscillation occurs very rapidly with a period of about 350~$\mathrm{fs}^{-1}$. The oscillation amplitude is proportional to 
$\sin{\phi_s}$. However it is damped by the LHCb decay time resolution, hence the relative resolution precision directly affects the relative precision on $\phi_s$. 
For an accurate determination of the resolution, prompt $J/\psi$ data is studied. The calibration yiels a decay time resolution of
$\sim 45\ \mathrm{fs}^{-1}$. Another damping factor arises from the wrong tagging probability $\omega_{tag}$ of the signal $B^0_s$ flavour. Flavour tagging 
algorithms use the information from same-side and opposite-side particles with respect to the signal channel to determine an effective tagging power 
$\epsilon_{tag}D^2$, where $D = (1-2\omega)$. The separate taggers are optimised using simulation samples and calibrated on data using flavour specific control 
channels. The taggers are combined to obtain an overall effective tagging power of the signal channel of $4.73 \pm 0.34 \%$.

The weighted unbinned likelihood fit is performed using a signal-only Probability Density Function (PDF)~\cite{PDF}. The fit procedure takes into account decay time
and angular acceptances, decay time resolution, as well as flavour tagging. The decay-time acceptance is determined using the control channel 
$B^0_d \rightarrow J/\psi K^{*0}(\rightarrow K^+ \pi^-)$, because it has a well-known lifetime and similar kinematics. The angular acceptance is taken from simulation 
by determining normalisation weights for all separate polarisation final states. The final result yields $\phi_s = -0.083 \pm 0.041 \pm 0.006\
\mathrm{rad}$~\cite{phis}, where the first uncertainty is statistical and the second systematic. This is the most precise 
single measurement of $\phi_s$ to date and is in agreement with the SM predictions~\cite{ckm}.

\section{Combination}

Many different decay channels have previously been exploited by the LHCb collaboration to determine $\phi_s^{c\bar{c}s}$.
These are the Run 1 analyses of $B^0_s \rightarrow J/\psi K^+K^-$~\cite{kk1}, $B^0_s \rightarrow J/\psi \pi^+\pi^-$~\cite{pipi1}, $B^0_s \rightarrow J/\psi K^+K^-$ for 
the $K^+K^-$ invariant mass region above the $\phi(1020)$~\cite{highkk}, $B^0_s \rightarrow \psi (2S) \phi$~\cite{psi2s} and $B^0_s \rightarrow 
D_s^+D_s^-$~\cite{dsds}, and the Run 2 analysis of $B^0_s \rightarrow J/\psi \pi^+\pi^-$~\cite{pipi2}. For the combination, potential systematic correlations have to 
be taken into account, as the analyses occasionally overlap with datasets and methods used, e.g. prompt $J/\psi$ data in the decay time resolution determination and 
$B^0_d \rightarrow J/\psi K^{*0}$ in the decay time efficiency determination. The combined LHCb result yields $\phi_s = -0.041 \pm 0.025\ \mathrm{rad}$~\cite{phis}, 
which is in agreement with the SM.

\begin{figure}[h]
\center
\includegraphics[scale=0.1]{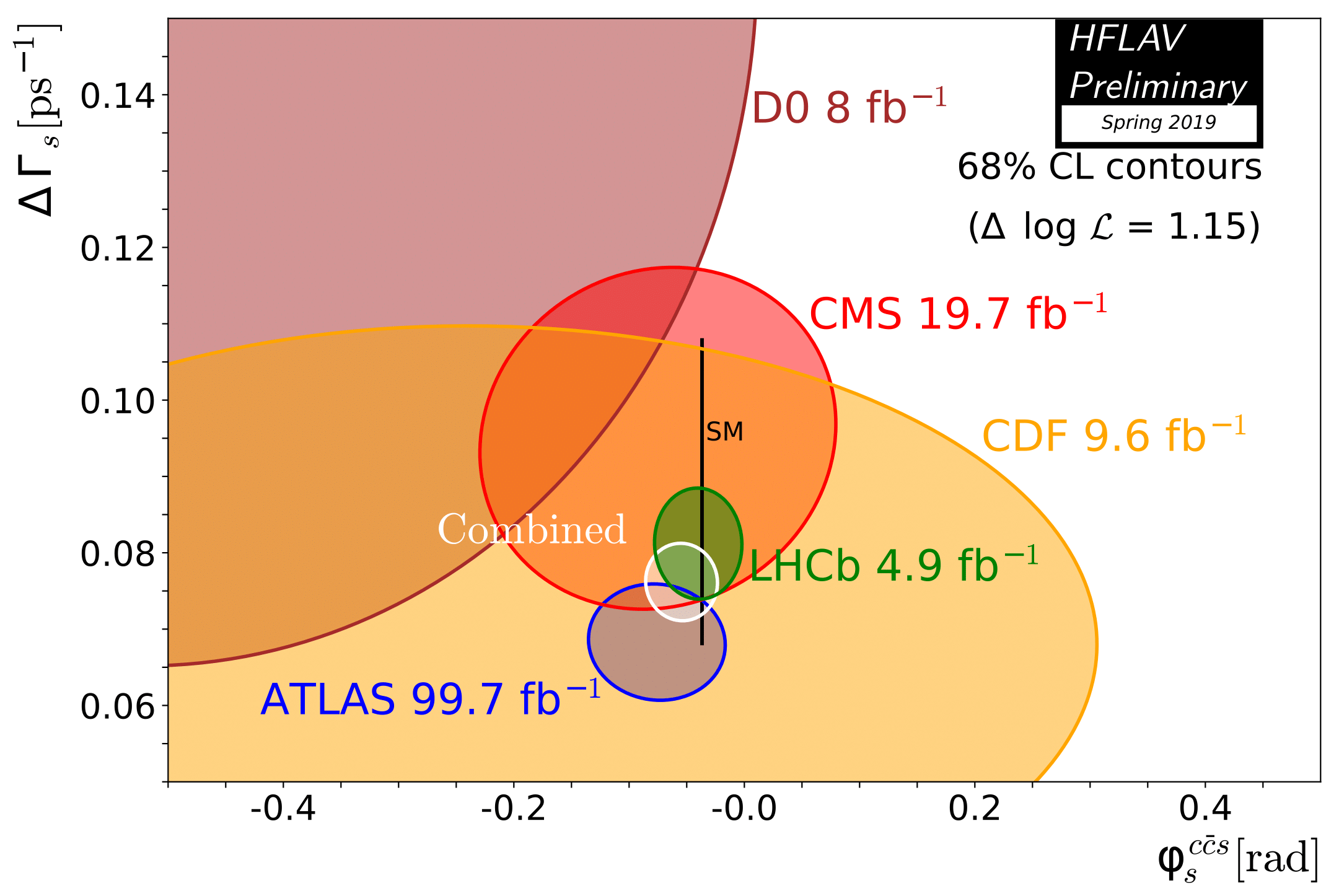}
\caption[]{68\% confidence level regions in $\phi_s$ vs $\Delta\Gamma_s$ plane obtained from individual contours of D0, CDF, ATLAS, CMS and LHCb measurements
	and the combined contour~\cite{HFLAV}. The prediction assuming the SM~\cite{ckm} is shown as a black thin rectangle.}
\label{fig:hflav}
\end{figure}

This result is further combined with measurements performed by Tevatron and further LHC experiments, namely the analysis of $B^0_s \rightarrow J/\psi \phi$ performed 
by the D0~\cite{D0}, CDF~\cite{CDF}, ATLAS~\cite{ATLAS} and CMS~\cite{CMS} collaborations. Figure~\ref{fig:hflav} shows the world average result determined by the 
Heavy Flavour Averaging Group. The value of the {\it CP} violating phase $\phi_s = -0.055 \pm 0.021\ \mathrm{rad}$~\cite{HFLAV}(Preliminary) is dominated by LHCb and 
in agreement with the SM. However the experimental precision is still significantly larger than the theoretical uncertainty, thus leaving enough room for New Physics.
More Run 2 data will be added to improve the statistical uncertainty, as well as data from future experimental upgrades with 
increased LHC luminosity.

\end{document}